\documentclass[mathleft,fleqn,%
]{an}
\usepackage{graphicx}
\usepackage[varg]{txfonts}
\newcommand\rfe[1]{(\ref{eq:#1})}
\newcommand\lab[1]{\label{eq:#1}}
\newcommand\nonu{\nonumber}
\newcommand\br{\begin{eqnarray}}
\newcommand\er{\end{eqnarray}}
\newcommand\be{\begin{equation}}
\newcommand\ee{\end{equation}}

\newcommand\foot[1]{\footnotemark\footnotetext{#1}}
\newcommand\lb{\lbrack}
\newcommand\rb{\rbrack}

\renewcommand\({\left(}
\renewcommand\){\right)}

\newcommand\bc{\begin{center}}
\newcommand\ec{\end{center}}

\newcommand\partder[2]{\frac{{\partial {#1}}}{{\partial {#2}}}}

\renewcommand\a{\alpha}

\renewcommand\d{\delta}

\newcommand\eps{\epsilon}
\newcommand\vareps{\varepsilon}

\newcommand\G{\Gamma}

\newcommand\h{\frac{1}{2}}
\renewcommand\k{\kappa}
\renewcommand\l{\lambda}

\newcommand\m{\mu}
\newcommand\n{\nu}
\newcommand\om{\omega}

\newcommand\vp{\varphi}
\renewcommand\P{\Phi}
\newcommand\pa{\partial}

\renewcommand\th{\theta}

\newcommand\vpdot{\stackrel{.}{\varphi}}

\newcommand\adot{\stackrel{.}{a}}
\newcommand\addot{\stackrel{..}{a}}

\begin{document}

\Pagespan{1}{}
\Yearpublication{2015}%
\Yearsubmission{2015}%
\Month{0}%
\Volume{999}%
\Issue{0}%
\DOI{asna.201500000}%

\title{Stable Emergent Universe -- A Creation without Big-Bang}

\author{Eduardo Guendelman\inst{1}\fnmsep\thanks{Corresponding author:
        {guendel@bgu.ac.il}}
\and Ram\'on Herrera\inst{2} \and  Pedro Labrana\inst{3}
\and Emil Nissimov\inst{4} \and Svetlana Pacheva\inst{4} 
}
\titlerunning{Emergent Universe}
\authorrunning{E. Guendelman, R. Herrera, P. Labrana, E. Nissimov, S.Pacheva}
\institute{
Department of Physics, Ben-Gurion University of the Negev, Beer-Sheva, Israel
\and 
Instituto de F\'{\i}sica, Pontificia Universidad Cat\'{o}lica de
Valpara\'{\i}so,  Avenida Brasil 2950, Casilla 4059,
Valpara\'{\i}so, Chile
\and 
Departamento de F\'{\i}sica, Facultad de Ciencias, 
Universidad del B\'{\i}o-B\'{\i}o, Casilla 5-C, Concepci\'{o}n, Chile
\and
Institute for Nuclear Research and Nuclear Energy,
Bulgarian Academy of Sciences, Sofia, Bulgaria
}

\received{XXXX}
\accepted{XXXX}
\publonline{XXXX}


\abstract{%
Based on an earlier introduced new class of generalized gravity-matter models defined 
in terms of two independent non-Riemannian volume forms (alternative generally 
covariant integration measure densities) on the space-time manifold, we derive an
effective ``Einstein-frame'' theory featuring the following remarkable
properties: (i) We obtain effective potential for the cosmological scalar
field possessing two infinitely large flat regions which allows for 
a unified description of both early universe inflation as well as of present dark
energy epoch; (ii) For a specific parameter range the model possesses a
non-singular {\em stable} ``emergent universe'' solution which describes an 
initial phase of evolution that precedes the inflationary phase.}

\maketitle

\section{Introduction}
Here we present a unified cosmological scenario of ``k-essence'' type 
(Chiba \textsl{et.al.} 2000; Armendariz-Picon \textsl{et.al.} 2000, 2001; Chiba 2002)
where both an inflation phase of the ``early'' universe and a slowly accelerated 
phase of the ``late'' universe do appear naturally from the existence of two 
infinitely large {\em flat regions} in the effective potential of the
pertinent cosmological scalar field, which we derive systematically from a 
well-defined Lagrangian action principle. Our starting point is the earlier proposed 
(Guendelman \textsl{et.al.} 2015a, 2015b) new kind of globally Weyl-scale invariant
gravity-matter action within the first-order (Palatini) approach formulated in 
terms of two different non-Riemannian volume forms (integration measures on
the spacetime manifold). The latter are constructed in terms of auxiliary maximal rank
antisymmetric tensor gauge fields called ``measure gauge fields''.
The cosmological scalar field has kinetic terms coupled to 
both non-Riemannian measures, and in addition to the scalar curvature 
term $R$ also an $R^2$ term is included (which is similarly allowed by global 
Weyl-scale invariance). Scale invariance is spontaneously broken upon solving 
the equations of motion for the auxiliary measure gauge fields 
due to the appearance of two arbitrary dimensionful integration constants. 

In the physical Einstein frame we obtain an effective k-essence 
(Chiba \textsl{et.al.} 2000; Armendariz-Picon \textsl{et.al.} 2000, 2001; Chiba 2002)
type of theory, where the effective scalar field potential has {\em two
infinitely-large flat regions}. The latter correspond to the two accelerating 
phases of the universe -- the inflationary early universe and the present 
``late''universe.

Another remarkable result we obtain within the flat region of the effective
scalar potential corresponding to the early universe is the appearance of an 
additional phase that precedes the inflation and describes a non-singular
{\em no Big Bang} creation of the universe. It is of an ``emergent universe'' type 
(Ellis \& Maartens 2004; Ellis \textsl{et.al.} 2004; Mulryne \textsl{et.al.} 2005) 
\textsl{i.e.}, the universe starts as a static Einstein universe, 
the scalar field rolls with a constant speed through a flat region and there is a 
domain in the parameter space of the theory where such non-singular solution exists 
and is {\em stable}. 

Concluding the introductory remarks let us point out that the formalism
employing alternative non-Riemannian volume forms in (generalized) gravity
triggers a number of physically interesting phenomena in spite of the
``pure-gauge'' nature of the auxiliary measure gauge fields.  
Apart from a new type of ``quintessential inflation'' scenario in cosmology describing
both the ``early'' and ``late'' universe in terms of a single scalar field and 
the uncovery of a stable initial non-singular ``emergent'' universe evolutionary phase 
(
Guendelman \textsl{et.al.} 2015a, 2015b; and here below) we have:
(i) new generic mechanism of dynamical generation of cosmological constant;
(ii) new mechanism of dynamical spontaneous breakdown of supersymmetry
in supergravity 
(Guendelman \textsl{et.al.} 2014, 2015c);
(iii) Coupling of non-Riemannian volume-form gravity-matter theories to
a special non-standard kind of nonlinear gauge system containing the
square-root of standard Maxwell/Yang-Mills Lagrangian yields charge
confinement/deconfinement phases associated with gravitational electrovacuum ``bag'' 
(Guendelman \textsl{et.al.} 2015d).

\section{Generalized Gravity-Matter Models Built With Two Independent Non-Riemannian
Volume-Forms}

Our starting point is a generalized modified-measure gravity-matter theory 
constructed in terms of two different non-Riemannian volume-forms (employing 
first-order Palatini formalism, and using units where $G_{\rm Newton} = 1/16\pi$)
(Guendelman \textsl{et.al.} 2015a, 2015b):
\br
S = \int d^4 x\,\P_1 (A) \Bigl\lb R + L^{(1)} \Bigr\rb
\nonu \\
+ \int d^4 x\,\P_2 (B) \Bigl\lb L^{(2)} + \eps R^2 + 
\frac{\P (H)}{\sqrt{-g}}\Bigr\rb \; .
\lab{TMMT}
\er
Here and below the following notations are used:
\begin{itemize}
\item
$\P_{1}(A)$ and $\P_2 (B)$ are two independent non-Riemannian volume-forms:
\be
\P_1 (A) = \frac{1}{3!}\vareps^{\m\n\k\l} \pa_\m A_{\n\k\l} \;\; ,\;\;
\P_2 (B) = \frac{1}{3!}\vareps^{\m\n\k\l} \pa_\m B_{\n\k\l} \; .
\lab{Phi-1-2}
\ee
\item
$\P (H)$ is the dual field-strength of a third auxiliary gauge field $H_{\m\n\l}$:
\be
\P (H) = \frac{1}{3!}\vareps^{\m\n\k\l} \pa_\m H_{\n\k\l} \; ,
\lab{Phi-H}
\ee
whose presence is essential for the consistency of \rfe{TMMT}.
\item
$R = g^{\m\n} R_{\m\n}(\G)$ and $R_{\m\n}(\G)$ are the scalar curvature and the 
Ricci tensor in the first-order (Palatini) formalism, where the affine
connection $\G^\m_{\n\l}$ is \textsl{a priori} independent of the metric $g_{\m\n}$.
In the second action term in \rfe{TMMT} we have added a $R^2$ gravity term
(again in the Palatini form)\foot{The gravity model $R+R^2$ within the
second order formalism was the first inflationary model originally
proposed in 
Ref.(Starobinsky 1980).}.
\item
$L^{(1,2)}$ denote two different Lagrangians of a single scalar matter field 
(``dilaton'' or ``inflaton'') of the form:
\br
L^{(1)} = -\h g^{\m\n} \pa_\m \vp \pa_\n \vp - V(\vp) \;\; ,\;\;
V(\vp) = f_1 e^{-\a\vp} \; ,
\lab{L-1} \\
L^{(2)} = -\frac{b}{2} e^{-\a\vp} g^{\m\n} \pa_\m \vp \pa_\n \vp + U(\vp) 
\;\; ,\;\; U(\vp) = f_2 e^{-2\a\vp} \; ,
\lab{L-2}
\er
where $\a, f_1, f_2$ are dimensionful positive parameters, whereas $b$ is a
dimensionless one.
\end{itemize}

The action \rfe{TMMT} possesses a {\em global Weyl-scale invariance}:
\br
g_{\m\n} \to \l g_{\m\n} \;,\; \G^\m_{\n\l} \to \G^\m_{\n\l} \; ,\; 
\vp \to \vp + \frac{1}{\a}\ln\l \; ,
\lab{scale-transf} \\
A_{\m\n\k} \to \l A_{\m\n\k} \; ,\; B_{\m\n\k} \to \l^2 B_{\m\n\k} \; ,\; 
H_{\m\n\k} \to H_{\m\n\k} \; .
\nonu
\er

The equations of motion w.r.t. affine connection $\G^\m_{\n\l}$ resulting
from the action \rfe{TMMT} yield the following solution for the latter:
\be
\G^\m_{\n\l} = \G^\m_{\n\l}({\bar g}) = 
\h {\bar g}^{\m\k}\(\pa_\n {\bar g}_{\l\k} + \pa_\l {\bar g}_{\n\k} 
- \pa_\k {\bar g}_{\n\l}\) \; ,
\lab{G-eq}
\ee
as a Levi-Civita connection corresponding to the Weyl-rescaled metric 
${\bar g}_{\m\n}$:
\be
{\bar g}_{\m\n} = (\chi_1 + 2\eps \chi_2 R) g_{\m\n} \;\; ,\;\; 
\chi_1 \equiv \frac{\P_1 (A)}{\sqrt{-g}} \;\; ,\;\;
\chi_2 \equiv \frac{\P_2 (B)}{\sqrt{-g}} \; .
\lab{bar-g}
\ee
Transition from the original metric $g_{\m\n}$ to ${\bar g}_{\m\n}$ accomplishes
the passage to the physical {\em ``Einstein-frame''}, where the gravity equations
of motion acquire the standard Einstein's form
$~R_{\m\n}({\bar g}) - \h {\bar g}_{\m\n} R({\bar g}) = \h T^{\rm eff}_{\m\n}$
with an appropriate {\em effective matter energy-momentum tensor} defined in terms
of an {\em effective Einstein-frame matter Lagrangian} $L_{\rm eff}$ 
(see \rfe{L-eff-final} below).

Variation of the action \rfe{TMMT} w.r.t. auxiliary tensor gauge fields
$A_{\m\n\l}$, $B_{\m\n\l}$ and $H_{\m\n\l}$ yields the equations:
\br
\pa_\m \Bigl\lb R + L^{(1)} \Bigr\rb = 0 \;, \;
\pa_\m \Bigl\lb L^{(2)} + \eps R^2 + \frac{\P (H)}{\sqrt{-g}}\Bigr\rb = 0 
\;, \; \pa_\m \Bigl(\frac{\P_2 (B)}{\sqrt{-g}}\Bigr) = 0 
\nonu \\
{}
\lab{A-B-H-eqs}
\er
whose solutions read:
\br
\frac{\P_2 (B)}{\sqrt{-g}} \equiv \chi_2 = {\rm const} \;\; ,\;\;
R + L^{(1)} = - M_1 = {\rm const} \; ,
\nonu \\
L^{(2)} + \eps R^2 + \frac{\P (H)}{\sqrt{-g}} = - M_2  = {\rm const} \; .
\lab{integr-const}
\er
Here $M_1$ and $M_2$ are arbitrary dimensionful and $\chi_2$
arbitrary dimensionless integration constants.

The first integration constant $\chi_2$ in \rfe{integr-const} preserves
global Weyl-scale invariance \rfe{scale-transf}
whereas the appearance of the second and third integration constants $M_1,\, M_2$
signifies {\em dynamical spontaneous breakdown} of global Weyl-scale invariance 
under \rfe{scale-transf} 
due to the scale non-invariant solutions (second and third ones) in \rfe{integr-const}. 

To elucidate the physical meaning of the three arbitrary integration constants 
$M_1,\, M_2,\,\chi_2$ we used in 
Refs.(Guendelman \textsl{et.al.} 2015b, 2014, 2015c)
the canonical Hamiltonian formalism. 
Namely, $M_1,\, M_2,\,\chi_2$ are identified as conserved Dirac-constrained
canonical momenta conjugated to the ``magnetic'' components of the auxiliary
maximal rank antisymmetric tensor gauge fields $A_{\m\n\l}, B_{\m\n\l}, H _{\m\n\l}$
entering the original non-Riemannian volume-form action \rfe{TMMT}. The rest
(``electric'') components of $A_{\m\n\l}, B_{\m\n\l}, H _{\m\n\l}$ appear
as Lagrange multipliers for the above Dirac constraints.

Varying \rfe{TMMT} w.r.t. $g_{\m\n}$ and using relations 
\rfe{bar-g}, \rfe{integr-const} we arrive at the standard form of Einstein 
equations for the rescaled  metric ${\bar g}_{\m\n}$, 
\textsl{i.e.}, the ``Einstein-frame'' equations:
\be
R_{\m\n}({\bar g}) - \h {\bar g}_{\m\n} R({\bar g}) = \h T^{\rm eff}_{\m\n}
\lab{eff-einstein-eqs}
\ee
with energy-momentum tensor corresponding according to the standard definition:
\be
T^{\rm eff}_{\m\n} = g_{\m\n} L_{\rm eff} - 2 \partder{}{g^{\m\n}} L_{\rm eff}
\lab{T-eff}
\ee
to the following effective (Einstein-frame) scalar field Lagrangian 
of non-canonical ``k-essence'' (kinetic quintessence) type 
(Chiba \textsl{et.al.} 2000; Armendariz-Picon \textsl{et.al.} 2000, 2001; Chiba 2002)
(here $X \equiv - \h {\bar g}^{\m\n} \pa_\m \vp \pa_\n \vp$ denotes 
the scalar kinetic term):
\be
L_{\rm eff} = A(\vp) X + B(\vp) X^2 - U_{\rm eff}(\vp) \; ,
\lab{L-eff-final}
\ee
where (recall $V=f_1 e^{-\a\vp}$ and $U=f_2 e^{-2\a\vp}$):
\br
A(\vp) \equiv 1 \!+\! \Bigl\lb \h b e^{-\a\vp} \! -\! \eps (V - M_1)\Bigr\rb
\frac{V - M_1}{U + M_2 + \eps (V - M_1)^2} \phantom{aa}
\lab{A-def} \\
B(\vp) \equiv \chi_2 \frac{\eps\Bigl\lb U + M_2 + (V - M_1) b e^{-\a\vp}\Bigr\rb
- \frac{1}{4} b^2 e^{-2\a\vp}}{U + M_2 + \eps (V - M_1)^2} \; , \phantom{aaa}
\lab{B-def}\\
U_{\rm eff} (\vp) \equiv 
\frac{(V - M_1)^2}{4\chi_2 \Bigl\lb U + M_2 + \eps (V - M_1)^2\Bigr\rb} \; .
\phantom{aaa}
\lab{U-eff}
\er

\section{Infinitely Large Flat Regions of the Effective Scalar Potential}

The effective scalar potential $U_{\rm eff}(\vp)$ \rfe{U-eff} possesses the
following remarkable feature -- the existence of two 
{\em infinitely large flat regions} as function of $\vp$ which is 
an {\em explicit realization of quintessential inflation scenario}
(Peebles \& Vilenkin 1999; Appleby \textsl{et.al.} 2010).

The explicit form of the two flat regions is as follows: 
\begin{itemize}
\item
{\em (-) flat region} -- for large negative values of $\vp$:
\be
U_{\rm eff}(\vp) \simeq U_{(-)} \equiv 
\frac{f_1^2/f_2}{4\chi_2 (1+\eps f_1^2/f_2)} \; ,
\lab{U-minus} 
\ee
\item
{\rm (+) flat region} -- for large positive values of $\vp$:
\be
U_{\rm eff}(\vp) \simeq U_{(+)} \equiv 
\frac{M_1^2/M_2}{4\chi_2 (1+\eps M_1^2/M_2)} \; ,
\lab{U-plus}
\ee
\end{itemize}

The qualitative shape of $U_{\rm eff}(\vp)$ \rfe{U-eff} is depicted on
Figs.1 and 2.

\begin{figure}
\begin{center}
\includegraphics[width=8cm,keepaspectratio=true]{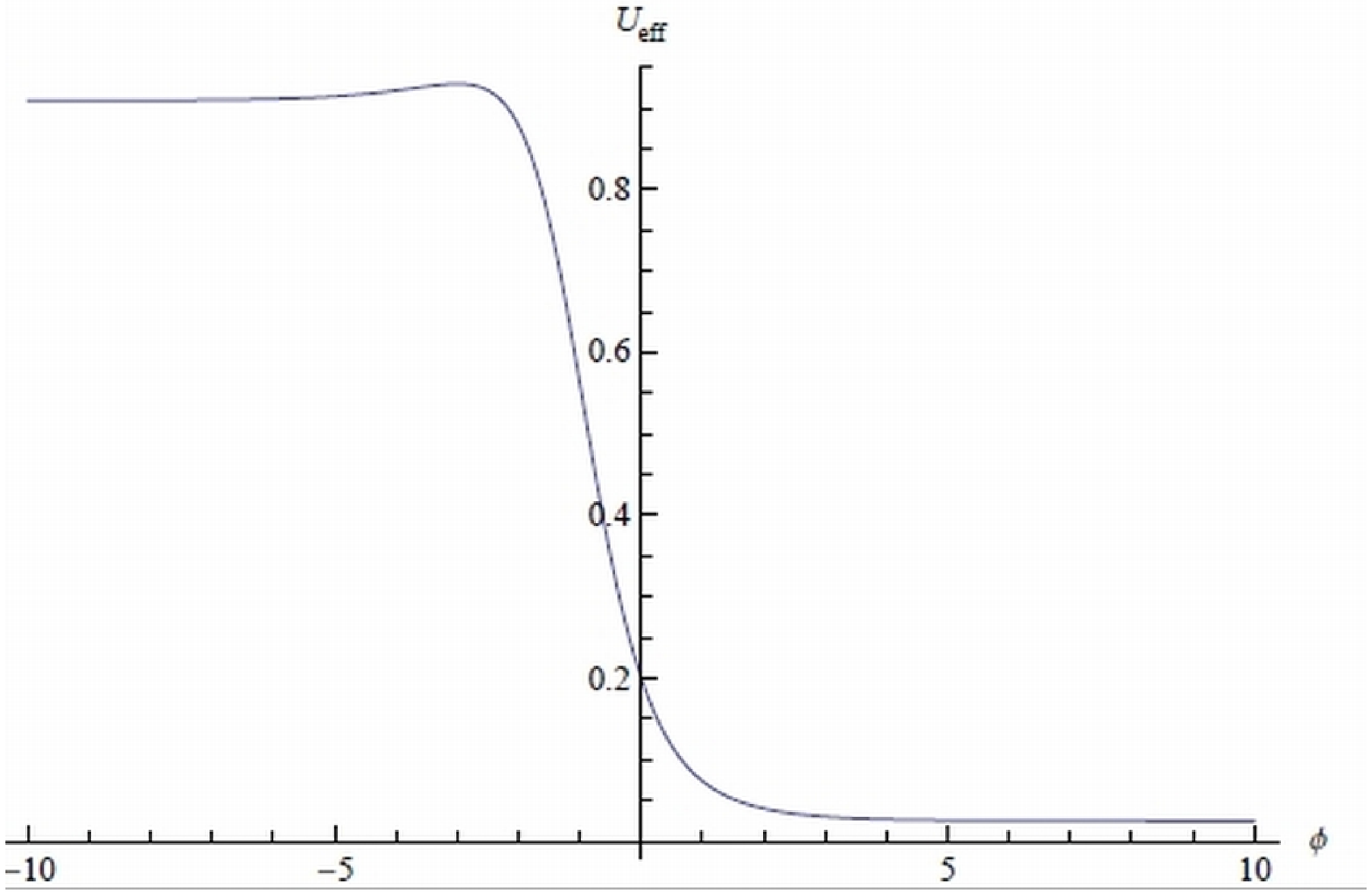}
\caption{Qualitative shape of the effective scalar potential $U_{\rm eff}$ \rf{U-eff}
as function of $\vp$ for $M_1 < 0$.}
\end{center}
\end{figure}

\begin{figure}
\begin{center}
\includegraphics[width=8cm,keepaspectratio=true]{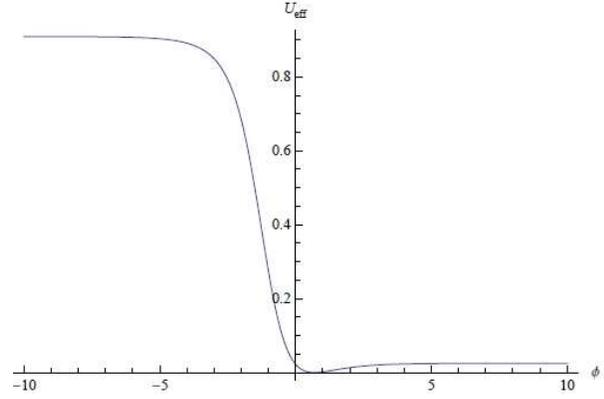}
\caption{Qualitative shape of the effective scalar potential $U_{\rm eff}$
\rfe{U-eff} as function of $\vp$ for $M_1 > 0$.}
\end{center}
\end{figure}

The flat regions \rfe{U-minus} and \rfe{U-plus} correspond 
to the evolution of {\em early} and the {\em late} universe, respectively, 
provided we choose the ratio of the coupling constants in the original scalar 
potentials versus the ratio of the scale-symmetry breaking integration constants to
obey the following strong inequality:
\be
\frac{f_1^2}{f_2} \gg \frac{M_1^2}{M_2} \quad ,\quad |\eps| \frac{M_1^2}{M_2} \ll 1 \; .
\lab{early-vs-late-2}
\ee
which makes the {\em vacuum energy density of the early universe} $U_{(-)}$ 
\rfe{U-minus} much bigger than that of the late universe $U_{(+)}$ \rfe{U-plus}).

If we choose the scales $|M_1| \sim M^4_{EW}$ and $M_2 \sim M^4_{Pl}$ 
(Arkani-Hamed \textsl{et.al.} 2000), where $M_{EW},\, M_{Pl}$ are the electroweak 
and Planck scales, respectively, we are then naturally led to a very small vacuum 
energy density:
\be
U_{(+)}\sim M^8_{EW}/M^4_{Pl} \sim 10^{-120} M^4_{Pl} \; ,
\lab{U-plus-magnitude}
\ee
which is the right order of magnitude for the present epoch's vacuum energy density.

On the other hand, if we take the order of magnitude of the coupling constants 
in the effective potential $f_1 \sim f_2 \sim (10^{-2} M_{Pl})^4$, then the order of
magnitude of the vacuum energy density of the early universe becomes:
\be
U_{(-)} \sim f_1^2/f_2 \sim 10^{-8} M_{Pl}^4 \; ,
\lab{U-minus-magnitude}
\ee
which conforms to the Planck Collaboration data 
(Adam \textsl{et.al.} 2015)
implying the energy scale of inflation to be of order $10^{-2} M_{Pl}$.

\section{Non-Singular Stable ``Emergent Universe'' Solution}

We start with the Friedman equations,  
see \textsl{e.g.} (Weinberg 1972):
\be
\frac{\addot}{a}= - \frac{1}{12} (\rho + 3p) \quad ,\quad
H^2 + \frac{K}{a^2} = \frac{1}{6}\rho \quad ,\;\; H\equiv \frac{\adot}{a} \; ,
\lab{friedman-eqs}
\ee
describing the universe' evolution. In the present case with ``Einstein
frame'' effective scalar field action \rfe{L-eff-final} the energy density $\rho$
and the pressure $p$ of the scalar field $\vp = \vp (t)$ read explicitly:
\br
\rho = \h A(\vp) \vpdot^2 + \frac{3}{4} B(\vp) \vpdot^4 + U_{\rm eff}(\vp) \; ,
\lab{rho-def} \\
p = \h A(\vp) \vpdot^2 + \frac{1}{4} B(\vp) \vpdot^4 - U_{\rm eff}(\vp) \; .
\lab{p-def}
\er
$H$ is the Hubble parameter and $K$ denotes the Gaussian curvature of the spacial 
section in the Friedman-Lemaitre-Robertson-Walker metric, 
see \textsl{e.g.} (Weinberg 1972):
\be
ds^2 = - dt^2 + a^2(t) \Bigl\lb \frac{dr^2}{1-K r^2}
+ r^2 (d\th^2 + \sin^2\th d\phi^2)\Bigr\rb \; .
\lab{FLRW}
\ee

``Emergent universe'' is defined as a solution of the Friedman Eqs.\rfe{friedman-eqs}
subject to the condition on the Hubble parameter $H$:
\br
H=0 \;\; \to \;\; a(t) = a_0 = {\rm const} \,,\;\; \rho + 3p =0 \; ,
\nonu \\
\frac{K}{a_0^2} = \frac{1}{6}\rho ~(= {\rm const}) \; ,
\lab{emergent-cond}
\er
with $\rho$ and $p$ as in \rfe{rho-def}-\rfe{p-def}. Here $K=1$ (``Einstein universe'').

The ``emergent universe'' condition \rfe{emergent-cond} implies that the $\vp$-velocity
$\vpdot \equiv \vpdot_0$ is time-independent and satisfies the bi-quadratic 
algebraic equation:
\be
\frac{3}{2} B_{(-)}\vpdot_0^4 + 2 A_{(-)}\vpdot_0^2 - 2 U_{(-)} = 0 \; ,
\lab{vpdot-eq}
\ee
where $A_{(-)},\, B_{(-)},\, U_{(-)}$ are the limiting values on the $(-)$ flat region
of $A(\vp),\, B(\vp),\, U_{\rm eff}(\vp)$ \rfe{A-def}-\rfe{U-eff}.

The solution of Eq.\rfe{vpdot-eq} reads:
\be
\vpdot_0^2 = - \frac{2}{3B_{(-)}} \Bigl\lb A_{(-)} \mp
\sqrt{A_{(-)}^2 + 3 B_{(-)}U_{(-)}}\Bigr\rb \; .
\lab{vpdot-sol}
\ee
and, thus, the ``emergent universe'' is characterized with {\em finite initial} 
Friedman factor and density:
\be
a_0^2 = \frac{6K}{\rho_0} \quad ,\quad
\rho_0 = \h A_{(-)}\vpdot_0^2 + \frac{3}{4} B_{(-)}\vpdot_0^4 + U_{(-)} \; ,
\lab{emergent-univ}
\ee
with $\vpdot_0^2$ as in \rfe{vpdot-sol}.

Analysis of stability of the ``emergent universe'' solution \rfe{emergent-univ} 
yields a harmonic oscillator type equation for the perturbation of the
Friedman factor $\d a$:
\be
\d \addot + \om^2 \d a = 0 \; ,
\lab{stability-eq-0}
\ee
with a ``frequency'' squared:
\be
\om^2 \equiv \frac{2}{3}\rho_0\,\frac{\sqrt{A_{(-)}^2 + 3B_{(-)}U_{(-)}}}{A_{(-)} -
2\sqrt{A_{(-)}^2 + 3 B_{(-)}U_{(-)}}} \; .
\lab{stability-eq}
\ee
Thus, stability condition $\om^2 >0$ implies the following constraint on the coupling 
parameters:
\br
{\rm max} \Bigl\{-2\,,\, -8\bigl(1+3\eps f_1^2/f_2\bigr)
\Bigl\lb 1 - \sqrt{1 - \frac{1}{4\bigl(1+3\eps f_1^2/f_2\bigr)}}\Bigr\rb\Bigr\}
\nonu \\
< b\frac{f_1}{f_2} < -1  \; .
\lab{param-constr}
\er

Since the ratio $\frac{f_1^2}{f_2}$ proportional to the
height of the $(-)$ flat region of the effective scalar potential,
\textsl{i.e.}, the vacuum energy density in the early universe, must be
large (cf. \rfe{early-vs-late-2}), we find that the lower end of the interval in 
\rfe{param-constr} is very close to the upper end, \textsl{i.e.}, 
$b\frac{f_1}{f_2} \simeq -1$.

From Eqs.\rfe{vpdot-sol}-\rfe{emergent-univ} we obtain an inequality satisfied by the
initial energy density $\rho_0$ in the emergent universe:
\be
U_{(-)} < \rho_0 < 2U_{(-)} \; ,
\lab{rho-0}
\ee
which together with the estimate of the order of magnitude for 
$U_{(-)}$ \rfe{U-minus-magnitude} implies order of magnitude for the initial
Friedman factor:
\be
a_0^2 \sim 10^{-8} K M_{Pl}^{-2} 
\lab{a-0}
\ee
($K$ being the Gaussian curvature of the spacial section).

\section{Concluding Remarks}

In Ref.(Guendelman \textsl{et.al.} 2015) 
the implications resulting from the present model for the ratio 
$r$ of tensor-to-scalar perturbations were studied. It was found that very
small values of the coupling parameter $\a$ (appearing in the initial scalar
potentials \rfe{L-1}-\rfe{L-2}) yield small values for $r$ (\textsl{e.g.}, the value 
$\a \simeq 10^{-20}$ correspond to $r \simeq 0.017$) which are well
supported by Planck data.

Furthermore, in Ref.(Guendelman \textsl{et.al.} 2015) 
the system of evolutionary equations
(the Friedman ones \rfe{friedman-eqs} plus the scalar field equations of 
motion resulting from the effective ``k-essence'' Lagrangian \rfe{L-eff-final})
was studied in some detail using the methods of autonomous dynamical systems.
A numerical evidence was found for the existence of a short transitional
phase of ``super-inflation'' (Labrana 2013) 
connecting the ``emergent'' and the ``slow-roll'' inflationary phases.

Few additional questions can be studied, for example the problem of reheating,
which one may worry about. The reason is that the effective scalar field
(inflaton) potential may either lack a minimum (Fig.1 above) or the pertinent 
minimum may be too shallow (Fig.2 above) so that this appears to imply the absence of 
an oscillatory behavior for the standard reheating scenario. 
It is possible to introduce a curvaton field, which takes care of reheating and 
primordial perturbations -- this can be done in a scale-invariant way 
(Guendelman \& Ramon 2015).

Another interesting subject concerns the quantum stability, extending the proof of 
classical stability, of the ``emergent universe'' solution to the quantum regime 
(del Campo \textsl{et.al} 2015). 

To recapitulate, let us list the main features illustrating the impact of
non-Riemannian volume-forms in generally-covariant theories:

\begin{itemize}
\item
Non-Riemannian volume-form formalism in gravity/matter theories 
(\textsl{i.e.}, employing alternative non-Riemannian reparametrization covariant 
integration measure densities on the spacetime manifold) naturally generates a 
{\em dynamical cosmological constant} as an arbitrary dimensionful 
integration constant.
\item
Employing two different non-Riemannian volume-forms leads to the construction of a
new class of gravity-matter models, which produce an effective scalar potential with 
{\em two infinitely large flat regions}. This allows for a unified description of both 
early universe inflation as well as of present dark energy epoch.
\item
A remarkable feature is the existence of a stable initial phase of
{\em non-singular} universe creation preceding the inflationary phase
-- ``emergent universe'' without ``Big-Bang''.
\end{itemize}

Further very interesting features of gravity-matter theories built with
non-Riemannian spacetime volume-forms include:
\begin{itemize}
\item
Within non-Riemannian-modified-measure minimal $N=1$ supergravity the 
dynamically generated cosmological constant triggers spontaneous supersymmetry
breaking and mass generation for the gravitino, \textsl{i.e.}, supersymmetric 
Brout-Englert-Higgs effect 
(Guendelman \textsl{et.al.} 2014, 2015c). Applying the same non-Riemannian
volume-form formalism to anti-de Sitter supergravity allows to produce
simultaneously a very large physical gravitino mass and a very small 
{\em positive} observable cosmological constant 
(Guendelman \textsl{et.al.} 2014, 2015c) in accordance with modern 
cosmological scenarios for slowly expanding universe of the present epoch
(Riess \textsl{et.al.} 1998,2004; Perlmutter \textsl{et.al.} 1999).
\item
Adding interaction with a special nonlinear (``square-root'' Maxwell) gauge field 
(known to describe charge confinement in flat spacetime) produces various
phases with different strength of confinement and/or with deconfinement, 
as well as gravitational electrovacuum ``bags'' partially mimicking the properties 
of {\em MIT bags} and solitonic constituent quark models (for details, see 
Ref.(Guendelman \textsl{et.al.} 2015d)).
\end{itemize}

\begin{quote}
  \verb+\acknowledgements+
\end{quote}

E.G., E.N. and S.P. gratefully acknowledge support of our collaboration through the 
academic exchange agreement between the Ben-Gurion University in Beer-Sheva,
Israel, and the Bulgarian Academy of Sciences. 
R.H. was supported by Comisi\'on Nacional de Ciencias y
Tecnolog\'{\i}a of Chile through FONDECYT Grant 1130628 and DI-PUCV 123.724.
P.L. was supported by Direcci\'on de Investigaci\'on de la
Universidad del B\'{\i}o-B\'{\i}o through grants GI121407/VBC and 141407 3/R.
S.P. and E.N. have received partial support from European COST Actions
MP-1210 and MP-1405, respectively, as well as from Bulgarian NSF Grant DFNI-T02/6.


\end{document}